\def\a{$\alpha$}
\def\b{$\beta$}

\documentstyle[psfig,12pt,twoside,fleqn]{article}

\title{Special Analytical Solutions of the Schr\"odinger Equation
for 2 and 3 Electrons in a Magnetic Field and
 {\em ad hoc} Generalizations to \mbox{N particles}} 
\author{by M. Taut\\Institut f\"ur Festk\"orper und Werkstoff-
 Forschung Dresden\\
Postfach 270018\\ 01171 Dresden, Germany\\
email: m.taut@ifw-dresden.de }
\begin{document}
\maketitle

\begin{abstract}
We found that the two--dimensional
Schr\"odinger equation for 3 electrons in an homogeneous
magnetic field (perpendicular to the plane) and a parabolic
scalar confinement potential (frequency $\omega_0$) 
has exact analytical solutions in the
limit, where the expectation value of the
center of mass vector $\bf R$ is small compared with the average 
distance between the electrons.
These analytical solutions exist only
for certain discrete values of the effective frequency
$\tilde \omega=\sqrt{\omega_o^2 + ({ \omega_c \over 2} )^2}$.
Further, for finite external fields,
the total angular momenta must be $M_L=3 m$ with $m=integer$,
and spins have to be parallel. The analytically solvable states are always
cusp states, and take the components of higher Landau levels into account. 
These special
analytical solutions for 3 particles 
and the exact solutions for 2 particles \cite{Taut2e}
can be written in
an unified form.
The first set of solutions reads\\
$
\Phi = \prod _{i<k} ({\bf r}_i-{\bf r}_k)^m\;
p_{n,m}(|{\bf r}_i-{\bf r}_k|)\;\;
\mbox{exp}\biggl(-\frac{1}{2} \; \tilde \omega_{n,m} \sum_l {\bf r}_l^2 \biggr)
$\\
where $p_{n,m}(x)$ are certain finite 
polynomials and $\tilde \omega_{n,m}$ is the spectrum of 
the fields. 
The pair angular momentum $m$ has to be an odd integer and the integer 
$n$ defines
the number of terms in the polynomials.
For infinite solvable fields $\tilde \omega_1$
there is a second set of  the form\\
$
\Phi =  {\cal A}_a \;\prod _{i<k} ({\bf r}_i-{\bf r}_k)^{m_{ik}}\;\;
\mbox{exp}\biggl(-\frac{1}{2} \; \tilde \omega_1 \sum_l {\bf r}_l^2 \biggr)
$\\
where ${\cal A}_a$ is the antisymmetrizer and the pair angular momenta
$m_{ik}$
can all be different integers.
In both cases the first factor is a short-- hand with
the convention  ${\bf r}^m=r^{|m|} e^{im\alpha}$.
These formulae, when {\em ad hoc} generalized to  N coordinates,
can be discussed as an ansatz for the wave function of the N--particle system.
This ansatz fulfills the following demands: it is exact for two particles
and for 3 particles in
the limit of small $\bf R$ and for the solvable external fields,
and it is an eigenfuncton of the total
orbital angular momentum. The Laughlin functions are special cases
of this ansatz
for infinite solvable fields and equal pair-- angular-- momenta. 
\end{abstract}
PACS classification:  \\
 8.30.Vw (Quantum dots etc.), 
 73.30.Hm (Quantum Hall Effect)\\

\newpage

\section{Introduction}
Most work on correlated electron systems in a magnetic field 
(and a
parabolic confinement potential) has been done adopting the following methods:
 Finite particle number ($N<10$) and finite field systems are tackled either by 
{\em numerical} expansion of the wave functions in
antisymmetrized products of one-- particle functions  
\cite{Girvin}--\cite{McDonald}
or 
{\em analytical ad hoc} approaches 
in the high field limit, where only the lowest Landau level (LLL) contributes
\cite{Laughlin}--\cite{Jain-QD}.
Other main streams are to use the Chern--Simons transformation 
\cite{Halperin} \cite{MacDonald-2}
and hoping that the transformed wave function can be guessed or
approximated more easily, or to use models for the electron-- electron
interaction \cite{Haldane} \cite{MacDonald-2} (All the above references are
mostly reviews).
In this paper we are trying another approach: 
We are looking for {\em analytical} solutions for few electron systems
and trying to generalize them {\em ad hoc} to N particle systems.
We use the genuine Coulombic electron-- electron interaction and do {\em not}
restrict ourselves to the lowest Landau level (LLL).
In a previous paper \cite{Taut2e} it has been shown that for $N=2$
(with Coulomb interaction between the electrons) there
is a 'spectrum'
of discrete external  field values for which the Schr\"odinger equation
can be solved
exactly and analytically. As shown below, these solutions comprise the
Laughlin states for $N=2$ as special cases.
The questions to be addressed in this paper are the following: Do similar
exact solutions also exist for three electrons? If so, does the corresponding
field spectrum agree with the spectrum for $N=2$?
Is there any connection between the discrete field spectrum for solvability
and the discrete fields (for given particle density) observed in the 
Quantum Hall effect?
Are the Laughlin states still among these special states as special cases?
To answer one of the questions in advance: We did not find exact analytical 
solutions for $N=3$. However, if we consider 
the center of mass coordinate 
${\bf R} = \frac{1}{3}
\sum_{i=1}^3 {\bf r}_i$
versus inter--particle distance as an
small parameter and expand the Hamiltonian in a multi-pole series, 
the three-- electron-- system can be decomposed into 3 
pair problems which have similar analytical solutions as the
two-- electron-- system.
The center of mass vector vanishes exactly in the classical ground state.
Therefore, one should expect that our expansion works well
for weak external fields or for systems, where after a Chern--Simons 
transformation and  a proper mean field approximation
(for finite systems!) the effective field is weak.
Surprisingly, also for 3 particles in the small {\bf R} limit,
the Laughlin states, which belong to {\em infinite} fields,
 are among the analytically solvable solutions.\\

The plan of this 
paper is the following. 
Sect.2 gives a survey on the results of the exact solutions of the 
electron pair problem.
Because in this paper the three-- electron-- problem
is traced back to three pair problems, this seems to be 
helpful.
In Sect.3 we define an orthogonal transformation 
for the three-- electron-- problem
which contains the center of mass $\bf R$ as a parameter, and then we 
expand the transformed 
Hamiltonian into a multi-pole series in $\bf R$.
In Sect.4 it is shown that,
in zero order in $\bf R$, the transformed Schr\"odinger equation
can be solved exactly and analytically for a certain set of external fields and
total angular momenta. In Sect.5, the eigenfunctions for 2 and 3 
particles are written
in an unified form and {\em ad hoc} generalized to arbitrary particle number.
This expression is compared with the Laughlin and Jain states.
In Sect.6 it is shown that the analytically solvable states  are just
those states where a cusp appears in the energy versus total 
angular momentum curve.
The accuracy of the multi-pole expansion  is tested in Sect.7 by calculation of
the energy eigenvalues in first order 
perturbation theory in the dipole and the quadrupole term of the 
Hamiltonian.

\newpage

\section{Exact solutions for two electrons in an 
homogeneous magnetic field}
In this section we summarize the results of 
a previous paper \cite{Taut2e}  on the two-- electron-- problem and add some
important subsequent unpublished findings.
In particular, we add the asymptotic solutions to our former pattern,
which had not been given the due attention in \cite{Taut2e},
and incorporate the electron-- electron coupling constant 
$\beta$ explicitly, in order to
be able to investigate the behavior of the exact solutions in
varying the coupling strength.
Completing and reviewing the two-- electron-- problem is important, because 
in the present work the three-- electron-- problem is traced back to
three two-- electron-- problems.
It has been shown  in \cite{Taut2e} that 
the Schr\"odinger equation for two electrons in an homogeneous magnetic field
plus an external parabolic scalar potential
has exact analytical solutions for a certain infinite, but discrete set
of field values (hereafter referred to as 'solvable fields')
\footnote{If we speak of 'fields' without specification
to a special one, we mean the effective oscillator frequency
$\sqrt{\omega_0^2+(\frac{\omega_c}{2})^2}$, which is the relevant
parameter.}.
Except for the  asymptotic case of infinite external fields,
which is part of this pattern,
there is a one-- to-- one correspondence between exact solutions and
solvable fields.
Such solutions exist for singlet and triplet states as well as ground and
excited states.
A further qualitative feature  is that these solutions occur, whenever
a correlated state (with electron-- electron interaction included)
 is degenerate with
an uncorrelated one (without electron-- electron interaction) 
\cite{Taut-unpublished}.
Moreover, for each total spin and orbital angular momentum
quantum number as well as for
a given degree of excitation (ground state, first excited state, etc.),
there is an infinite series of solvable fields which converges
to zero. This means that
the solvable field values are dense at zero.\\
Now we are going to describe the general analytical form of the 
exact solutions. After introducing relative and center of mass  coordinate
\begin {equation}
{\bf r}={\bf r}_2-{\bf r}_1~~~~~;~~~~~{\bf R}={1\over 2}({\bf r}_1+{\bf
r}_2)
\end{equation}
the Hamiltonian (in atomic units $\hbar=m=e=1$) 
\begin{equation}
H=\sum\limits^2_{i=1}\biggl\{{1\over 2}\biggl({\bf p}_i+
{1\over c}{\bf A}({\bf r}_i)\biggl)^2 +
{1\over 2}\omega_o^2r_i^2\biggl\}+{\beta \over |{\bf r}_2-{\bf r}_1|}
+H_{spin}
\end{equation}
decouples exactly. 
\begin{equation}
H=2 \;H_r+{1\over 2} \;H_R+H_{spin}
\end{equation}

We follow the notation of \cite{Taut2e}
as long as not explicitly mentioned.  
 $\omega_o$ is the oscillator frequency of
 the parabolic external
confinement potential and ${\bf A}({\bf r})=\frac{1}{2}{\bf B}\times {\bf r}$
the vector potential of
the external magnetic field.
The center of mass
 degree of freedom behaves like a
quasi-particle in  rescaled external fields and the quasi-particle of the
relative coordinate is a particle in  rescaled external fields plus 
a rescaled repulsive Coulomb field 
originating from the \mbox{e-- e-- interaction}.
 The Schr\"odinger equation of the first problem
is trivial, the latter problem, which  will be considered here,
is described by the Hamiltonian
(see eq.(5) in \cite{Taut2e})
\begin {equation}
H_r={1\over 2}\biggl[{\bf p}+
{1\over c}{\bf A}_r\biggr]^2+{1\over
2}\omega_r^2r^2+\frac{\beta}{2r}
\end{equation}
where \footnote{The index '$r$'  and '$R$' refers to the relative 
and c.m. coordinate systems, respectively}
$\omega_r={1\over 2}\omega_o$, ${\bf A}_r={1\over 2}{\bf A}({\bf r})$.
In polar coordinates ${\bf r}=(r,\alpha)$,  the following ansatz
for the eigenfunction
\begin{equation}
\phi={e^{im\alpha}\over \sqrt{2\pi}}~~{u(r)\over
r^{1/2}}~~~~~;~~~~~m=0,\pm 1,\pm 2,\ldots
\label{ansatz-2e}
\end{equation}
is justified, where the Pauli principle demands that 
$m$ is even or odd in the singlet and triplet state, respectively.
The Schr\"odinger equation $H_r \; \phi({\bf r})=\epsilon_r \; \phi({\bf r})$
gives rise to the radial Schr\"odinger equation for $u(r)$
\begin{equation}
\biggl\{-{1\over 2}~{d^2\over dr^2}+{1\over 2}\biggl(m^2-{1\over 4}\biggr)
{1\over r^2}+{1\over 2}\tilde\omega_r^2 r^2+\frac{\beta}{2r}
\biggl\}u(r)=\tilde \epsilon_r \; u(r)
\label{rad-SGl}
\end{equation}
where $\tilde\omega_r={1\over 2}\tilde\omega=
{1\over 2}\sqrt{\omega_o^2 + ({ \omega_c \over 2} )^2}$ , 
$\tilde \epsilon_r={1\over 2}\tilde \epsilon=\epsilon_r-{1\over 4} m \omega_c$, 
$\omega_c={B \over c}$
and the solution is subject
to the normalization condition $\int\limits^\infty_o dr|u(r)|^2=1$.
In dimensionless variables $r \rightarrow \sqrt{ \tilde \omega_r} r$
and $\tilde \epsilon_r \rightarrow \tilde \epsilon_r /  \tilde \omega_r$  
the radial Schr\"odinger equation reads
\begin{eqnarray}
\biggl\{-{1\over 2}~{d^2\over d(\sqrt{ \tilde \omega_r} r)^2}+
{1\over 2}\biggl(m^2-{1\over 4}\biggr)
{1\over(\sqrt{ \tilde \omega_r} r)^2}+{1\over 2}(\sqrt{\tilde\omega_r} r)^2+
\frac{\beta}{\sqrt{ \tilde \omega_r}} \frac{1}{2 ( \sqrt{ \tilde \omega_r}  r)}
\biggl\}u(r)=  \nonumber \\
= \biggl( \frac{\tilde \epsilon_r}{\tilde \omega_r} \biggr) \; u(r)
\label{rad-SGl-dimless}
\end{eqnarray}
The {\em exactly solvable} eigenfunctions 
have the following form
\begin{equation}
u(r)=r^{|m|+\frac{1}{2}}\; p(r) \; e^{-\frac{1}{2}\; \tilde \omega_r \; r^2}
\label{u}
\end{equation}
where $p(r)$ is a {\em finite} polynomial
\begin{equation}
p(r)=\sum_{\nu=0}^{(n-1)} a_\nu \cdot (\sqrt{ \tilde \omega_r} r)^\nu
\end{equation}
with $n$ terms.
The soluble fields and the corresponding
eigenvalues are determined by the two requirements 
\begin{equation}
a_n=0~~~~~;~~~~~a_{n+1}=0 
\label{truncation}
\end{equation}
which guarantee truncation of the power series and therefore
normalizability of the eigenfunctions.
It is clear from (\ref{rad-SGl-dimless}) 
 that the two truncation conditions depend from 
the parameters only in the combination $\frac{\beta^2}{\tilde \omega_r}$ 
and $\frac{\tilde\epsilon_r}{\tilde\omega_r}$. Consequently, we have two
equations (\ref{truncation}) for effectively two parameters. 
Technically, we first calculate the solvable 
fields  $\frac{\beta^2}{\tilde\omega_r}$ 
from (16) in \cite{Taut2e} (with $\beta$ included) 
and then we get the corresponding eigenvalues
from (17) in \cite{Taut2e}, which we rewrite here in the form
\begin{equation}
\frac{\tilde\epsilon_r}{\tilde\omega_r}=  |m|+n
\label{eigen}
\end{equation}
Note that (\ref{eigen}) does not contain $\beta$ explicitly,
but only through the fact that  the solvable $\tilde\omega_r$ depend on beta.
The calculation of the solvable fields is non--trivial
 (and not repeated here in detail), but
when they are found, the calculation of the corresponding eigenvalues is 
simple through (\ref{eigen}).
Observe that the eigenvalues of (\ref{rad-SGl}) without e-e-interaction read
\cite{one-e-dot}
\begin{equation}
\frac{\tilde\epsilon_r}{\tilde\omega_r}=  |m|+2k+1
\label{eigen0}
\end{equation}
where $k$ is the node number. Comparison of (\ref{eigen}) and (\ref{eigen0})
shows the above mentioned degeneracies of the interacting system with the
noninteracting one. 
With $n=2k+1$,  the interaction-- free solutions fit 
into a generalized
pattern of all analytical solutions.
It can also be shown that equation (16) with (17) in \cite{Taut2e},
which defines the solvable 
$\frac{\beta^2}{\tilde\omega_r}$, 
has always the solution $\frac{\beta^2}{\tilde\omega_r}=0$ 
whenever $n$ is odd. As an example, see (20a) in \cite{Taut2e} for 
the case $n=3$.  (These infinite field solutions are not included in 
Tables 1 and 2 in \cite{Taut2e}.)
It is also clear from physical reasoning that for infinite external fields
the e--e--interaction has no influence on the eigensolutions, because the
kinetic energy dominates. Therefore, solutions without e--e--interaction
are exact for $\tilde \omega_r \rightarrow \infty$. 
Sometimes we shall call these solutions 'asymptotic' solutions.\\
The completed pattern  has the following over-- all-- structure: 
For $n=1,2$ there is
one solution (ground states), for $n=3,4$ we
have two solutions (one ground and one excited state), etc. (see Figure 1).
Generally, the soluble fields are the smaller
the larger the corresponding $n$ is. For $n \rightarrow \infty$ the
corresponding soluble $\tilde\omega_r$ converge to $0$.\\
\begin{figure}[!tbh]
\vspace{-2cm}
\begin{center}
{\psfig{figure=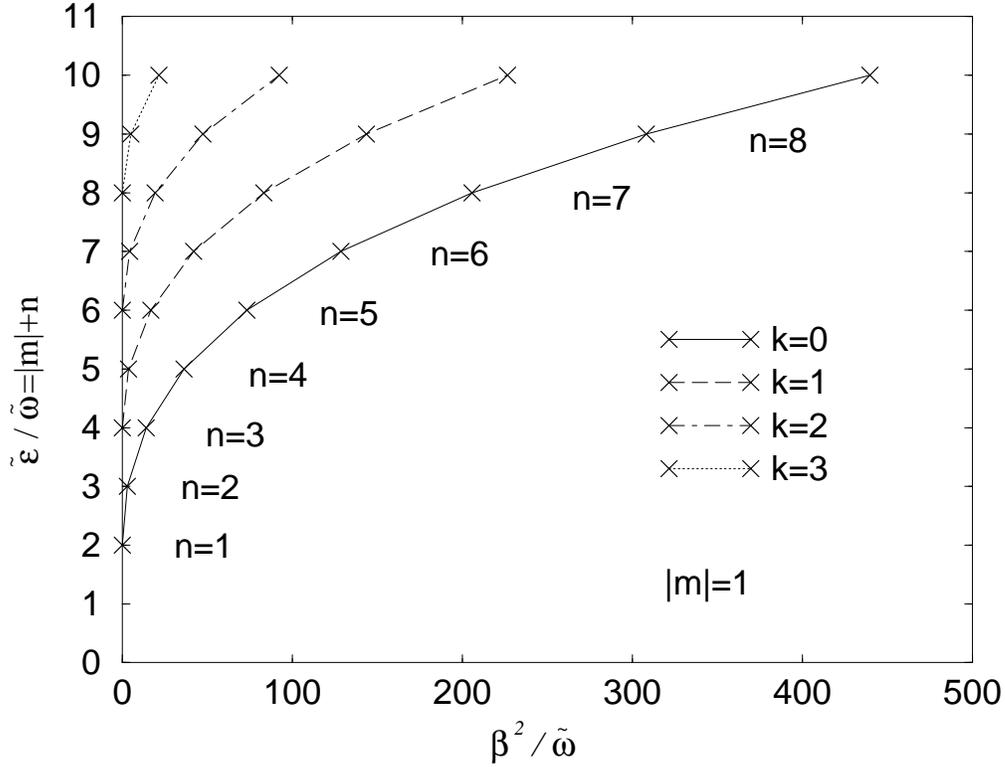,angle=-90,width=15.cm,bbllx=15pt,bblly=45pt,bburx=580pt,bbury=750pt}}
\caption[ ]{
Reduced energies (energy over effective oscillator frequency)
 versus squared coupling constant over
effective oscillator frequency for  relative angular momentum $|m|=1$.
The crosses indicate solvable states. The lines connect states with the
same node number $k$. The termination index $n$ is the same for all horizontal
rows of crosses with the same ordinate.
}
\label{fig1}
\end{center}
\end{figure}

Now we consider the eigenfunctions.
The series $a_\nu$ is defined by a 
two step recursion relation 
which reads for the soluble states 
(insert (17) into (14) in \cite{Taut2e})
\begin{eqnarray}
a_0&=&\mbox{normalization constant}\nonumber\\
a_1&=&\frac{1}{(2|m|+1)} \;\frac{\beta}{\sqrt{ \tilde \omega_r}}\;a_0 \nonumber\\
a_2&=&\frac{1}{4(|m|+1)} 
\biggl\{ \frac{1}{(2|m|+1)}\frac{\beta ^2}{ \tilde \omega_r} 
-2(n-1)  \biggr\} \; a_0 
\nonumber \\
\cdots  \nonumber \\
a_\nu&=&\frac{1}{\nu \;(\nu+2|m|)}
\biggl\{  \frac{\beta}{\sqrt{ \tilde \omega_r}}\;
a_{\nu-1}+2\;(\nu-n-1)\; a_{\nu-2} 
\biggr\}
\label{recursion}
\end{eqnarray}
It produces rather complicated expressions for larger $\nu$.
The recursion can also be started at the other end with $a_{n-1}$ as 
a normalization constant.\\
The eigenfunctions of asymptotic solutions 
fit also into the generalized scheme. 
For $\frac{\beta^2}{\tilde\omega_r}=0$, the recursion relation 
(\ref{recursion}) provides 
only non--vanishing coefficients with even index, 
meaning, that $p(r)$ is  a function
of $r^2$. If we insert $n=2k+1$, 
and $\nu=2 p$ with $p=0,1,2,...$ into (\ref{recursion}) we obtain the recursion
relation 
\begin{equation}
a_p=\frac{(p-k-1)}{p(p+|m|)}\;a_{p-1}
\end{equation}
which belongs to the
Generalized Laguerre polynomials. \footnote{We use the 
definition in Abramowicz, Stegun; Handbook of Mathematical Functions}
\begin{equation}
p_{n=2k+1,m}(r)=L_k^{|m|}(\tilde\omega_r\; r^2)
\end{equation}
Consequently, the 
Generalized Laguerre polynomials are a special case of our 
polynomials $p_{n,m}(r)$.
If we go in Fig.1  along the line for ground states ($k=0$) from the left 
to the right,
then the polynomials for the exactly solvable cases 
have the following form. For $n=1$ ,
$p_{n=1,m}(r)=L_0^{|m|}(\tilde\omega_r\; r^2)$ is 
a constant, for 
$n=2$ (simplest case with finite fields) 
$p_{n=2,m}(r)$ is a linear function without
a positive zero, for $n=3$ we have a quadratic function 
without positive zeros, etc.
Analogously, the exact solutions 
for the first excited state are all polynomials
with one node, but increasing order.\\

As an overview, we  give the formulae for the simplest exact solutions
of the completed pattern (see also (19) and (20) in \cite{Taut2e} with
$\beta$ included).
 For $n=1$ there is only the
asymptotic solution
\begin{eqnarray}
\frac{\beta^2}{\tilde \omega_r}&=&0 \label{om-N=2-n=1}\\
p(r)&=&1
\end{eqnarray}
For $n=2$ there is one finite-- field solution
\begin{eqnarray}
\frac{\beta^2}{\tilde\omega_r}&=&2 \; (2\; |m|+1) \label{om-N=2-n=2}\\
p(r) &=& \biggl[ 1+\frac{\beta \;r} {(2\; |m|+1)} \biggr]
\label{u-N=2-n=2}
\end{eqnarray}
Both former solutions are ground states.
For $n=3$ there is one asymptotic solution, which is a first excited state,
\begin{eqnarray}
\frac{\beta^2}{\tilde \omega_r}&=&0\\
p(r)&=&1-\frac{\tilde \omega_r\; r^2}{(|m|+1)}
\end{eqnarray}
and one at finite fields, which is a ground state,
\begin{eqnarray}
\frac{\beta^2}{\tilde\omega_r}&=&4\; (4\; |m|+3)
\label{om-N=2-n=3}\\
p(r) &=& \biggl[ 1+\frac{\beta \; r} {(2\; |m|+1)}
          +\frac{(\beta \; r) ^2}
                {2 (2\; |m|+1)(4\;|m|+3)} \biggr]
\label{u-N=2-n=3}
\end{eqnarray}
The corresponding energies follow from (\ref{eigen}), and
we put $a_0=1$ for the constant term of the polynomial $p(r)$ without loss
of generality.\\
In order to give an idea in what magnetic field range these exact
solutions are located we consider the case without confinement ($\omega_0=0$).
Then $\tilde\omega_r=\omega_c/4$, where all solvable frequencies 
given above
are in effective atomic units $a.u.^*$ ($\hbar=m^*=\beta=1$). 
On the other hand, for GaAs we have $B[Tesla]=\omega_c[a.u.^*]/0.1363$.
This means, that the {\em largest finite} solvable field ( for $n=2$ and $m=0$ )
is  $\omega_c=2\;a.u.^*$ and $B=14.7\; Tesla$. \\
 Now we add some words about the limit $\beta \rightarrow 0$. The 
wave functions of the asymptotic solutions
($\frac{\beta^2}{\tilde \omega_r}=0$) do not depend on $\beta$ at all,
indicating that they are robust against a variation of the e-- e-- interaction.
$\frac{\beta^2}{\tilde \omega_r}=0$ can be realized either by vanishing e-- e-- interaction ($\beta \rightarrow 0$) or infinite fields ($\tilde \omega_r
\rightarrow \infty$). On the other hand, 
the wave functions of the finite-- field-- solutions 
($\frac{\beta^2}{\tilde \omega_r}=finite$) 
evolve steadily into non-- interacting ones for $\beta \rightarrow 0$
indicating a strong dependence on the form of the e-- e-- interaction.\\

If we insert for the center of mass system the ground state WF, then the total
spatial WF  in the particle coordinates reads (use (\ref{ansatz-2e}),
(\ref{u}), and from \cite{Taut2e} formula (8))
\begin{equation}
\Phi({\bf r}_1,{\bf r}_2)=({\bf r}_1-{\bf r}_2)^m 
\;p_{n,m}(|{\bf r}_1-{\bf r}_2|)
\; e^{-\frac{\tilde\omega}{2} (r_1^2+r_2^2)}
\label{tot-WF-2e}
\end{equation}
where the first factor is a shorthand with the convention 
${\bf r}^m=r^{|m|} e^{im\alpha}$. In complex coordinates 
$ z=x+iy$, $\bar z=x-iy$, 
and with the opposite sign for the angular momentum 
\footnote{Observe, our $m$ is the angular momentum quantum number itself
 and negative
for the LLL.} $\bar m=-m $
(compared with Laughlins notation \cite{Laughlin} a 'bar' has been added), 
this means
\begin{eqnarray}
{\bf r}^m&=& z^m ~~~~~ \mbox{for} ~~~~~m\ge0\\
         &=& \bar z^{\bar m} ~~~~~ \mbox{for}  ~~~~~m\le0 
\end{eqnarray}
Therefore, for $m \le 0$, (\ref{tot-WF-2e}) reads in complex coordinates
\begin{equation}
\Phi(\bar z_1,\bar z_2)=(\bar z_1-\bar z_2)^{\bar m}
\;p_{n,\bar m}(|\bar z_1-\bar z_2|)
\; e^{-\frac{\tilde\omega}{2} (|\bar z_1|^2+|\bar z_2|^2)}
\label{WF-2e-complex}
\end{equation}
The solution for $n=1$ (infinite field, $p(x)=a_0=const.$)
agrees exactly with the Laughlin-- WF, in particular, $\bar m=1$ is
a determinant of two LLL functions, which corresponds to an uncorrelated 
full LLL.\\
For $m \ge 0$ we have
\begin{equation}
\Phi(z_1, z_2)=( z_1- z_2)^{m}
\;p_{n, m}(|z_1-z_2|)
\; e^{-\frac{\tilde\omega}{2} (|z_1|^2+|z_2|^2)}
\end{equation}
which is the complex conjugate of (\ref{WF-2e-complex}) 
and therefore the solution
for the opposite direction of the magnetic field, if $\omega_0=0$. 
In particular, $n=1$ and $m=1$ is a {\em  determinant} of the 
two one-- particle-- states
with $k=0$ and $m=0$ and $m=1$, in other words, it corresponds to 
the uncorrelated solution of two full LLs.
In summary, the solutions for $n=1$ comprise the Laughlin WFs and the
full LLs. It is tempting to assume that the other solutions 
($n=2,3,...$) have some
relation to the other Quantum Hall states, but until now there is no
prove for it. (see also Sect. 5)\\

\newpage

\section{Approximate decoupling for three electrons}
The Hamiltonian for three electrons in an homogeneous magnetic field
(vector potential ${\bf A}({\bf r})$) and a harmonic scalar potential
(oscillator frequency $\omega_0$) reads
\begin{equation}
H=\sum_{i=1}^3\biggl[{1\over 2}\biggl(\frac{1}{i}{\bf \nabla}_i
+{1\over c}{\bf
A}({\bf r}_i)\biggr)^2 +{1\over 2}\omega_o^2r_i^2\biggr]
+\sum_{i<k}{\beta \over |{\bf r}_i-{\bf r}_k|}+H_{spin}
\label{h-orig}
\end{equation}
where $H_{spin}=g\; \sum\limits_{i=1}^3 \;{\bf s}_i \; \cdot{\bf B}$.
The goal of the following considerations is the decoupling
of the Hamiltonian into a sum of independent Hamiltonians for quasi--
 particles.\\
For {\em two electrons}
 this happens automatically by introducing the center of mass
and relative coordinate. This is mainly due to the peculiarity
 that there is only 
one interaction term $\beta \over {|{\bf r}_2-{\bf r}_1|}$ which contains only
one of the new coordinates $\bf r$ and $\bf R$. 
The peculiarity for {\em  three electrons},  which can be taken advantage of,
  is the fact that the number of interaction terms
is equal to the number of particles.
This does not allow for an exact decoupling, but an approximate one into three
noninteracting pairs plus a coupling term which is small in the strong
correlation limit.\\
It is easily shown that an {\em orthogonal transformation}
\footnote{Exactly speaking, for keeping the one-- particle-- terms
of the Hamiltonian decoupled it suffices that the {\em row } vectors
of the matrix in (\ref{trafo})  are mutually orthogonal. 
However, using this more
general type of transformation does not provide any advantage in our case,
 because the additional freedom destroys the symmetry among the particles.}
 leaves the kinetic energy  in a
homogeneous magnetic field and the potential energy in an external harmonic
scalar potential, invariant. On the other hand, the center of mass (c.m.) of
a classical system in the ground state vanishes and 
it is natural to assume that the c.m. in the high correlation limit can
be treated as a small expansion parameter.
Therefore,
we look for an orthogonal transformation 
which transforms the e-- e-- interaction in such a form
that each term depends only on {\em one} of the new coordinates and the center
of mass. Additionally, we demand that the center of mass is invariant under 
the transformation, which  guarantees that if it is small in the original 
coordinates, so it is in the transformed ones.
The transformation from the original coordinates ${\bf r}_i$ to the
new ones ${\bf x}_i$, which fulfills all these requirements, reads
\begin{equation}
 \left[ \begin{array} {c} {\bf x}_1\\{\bf x}_2\\{\bf x}_3\end{array} \right]= 
\left[ \begin{array} {ccc}1/3&a&b\\b& 1/3&a\\a&b&1/3\end{array}\right]
\left[ \begin{array} {c} {\bf r}_1 \\ {\bf r}_2 \\ {\bf r}_3 \end{array} \right]
\label{trafo}
\end{equation}
where $a=1/3-1/\sqrt{3}$ and $b=1/3+1/\sqrt{3}$ and its inverse is
\begin{equation}
 \left[ \begin{array} {c}{\bf r}_1\\{\bf r}_2\\{\bf r}_3 \end{array} \right] =  
\left[ \begin{array} {ccc}1/3&b&a\\a& 1/3&b\\b&a&1/3\end{array}\right]
\left[ \begin{array} {c} {\bf x}_1 \\ {\bf x}_2 \\ {\bf x}_3 \end{array} \right]
\label{inverse-trafo}
\end{equation}
From (\ref{inverse-trafo}) it follows that
\begin{eqnarray}
{\bf r}_1-{\bf r}_2=\sqrt{3}\; \biggl({\bf X}-{\bf x}_3\biggr)
 \nonumber \\
{\bf r}_2-{\bf r}_3=\sqrt{3}\; \biggl({\bf X}-{\bf x}_1\biggr)
\label{diff}\\
{\bf r}_3-{\bf r}_1=\sqrt{3}\; \biggl({\bf X}-{\bf x}_2\biggr)
\nonumber 
\end{eqnarray}
so that the Hamiltonian in the new 
coordinates is
\begin{equation}
H=\sum_{i=1}^3\biggl[{1\over 2}\biggl(\frac{1}{i}{\bf \nabla}_i+{1\over c}{\bf
A}({\bf x}_i)\biggr)^2 +{1\over 2} \; \omega_o^2 \; x_i^2
+\frac{1}{\sqrt{3}} \; {\beta \over |{\bf x}_i-{\bf X}|} \biggr]
+H_{spin}
\label{h-trans}
\end{equation}
where  ${\bf X} \equiv \frac{1}{3} \sum_{i=1}^3 {\bf x}_i ={\bf R}$
 is the 
center of mass in the new coordinates.
It is possible (but complicated) to take care of
 the Pauli principle in the new coordinates. 
It is much easier first to
transform the wave functions (WF) back to the original coordinates and 
then do the antisymmetrization .  \\
While being still exact, (\ref{h-trans}) is not exactly 
decoupled because ${\bf X}$
contains all coordinates. For  ${\bf X}$ small compared with ${\bf x}_i$ , 
the e-e-interaction term 
can be expanded in a {\em multi-pole series }
\begin{equation}
V_{ee}=\sum_{l=0}^\infty V_{ee}^{(l)}
\end{equation}
where
\begin{eqnarray}
V_{ee}^{(0)}&=& \frac{\beta}{\sqrt{3}} \; \sum_{i=1}^3 \; \frac{1}{|{\bf x}_i|}
                                          \label{monopole}\\
V_{ee}^{(1)}&=& \frac{\beta}{\sqrt{3}} \; \sum_{i=1}^3 \; 
           \frac{{\bf X} \cdot {\bf x}_i}{|{\bf x}_i|^3} \label{dipole}\\
V_{ee}^{(2)}&=& \frac{\beta}{\sqrt{3}} \; \frac{1}{2}\; \sum_{i=1}^3 \; 
\biggl[ 3 \frac{({\bf X} \cdot {\bf x}_ i)^2} {|{\bf x}_i|^5}
        - \frac{({\bf X})^2 }{|{\bf x}_i|^3} \biggr] \label{quadrupole}\\
\cdots \nonumber
\end{eqnarray}
In zero order in ${\bf X}$, the Hamiltonian $H^{(0)}$ 
is decoupled and can be solved
exactly and in many cases even analytically (see below).
The general strategy for amending the zero order result
is to  consider the multi-pole terms
in perturbation theory.
As a response to a frequently asked questions, we want to emphasize 
the following.
The approximation ${\bf X}=0$ does not mean that the new
coordinates ${\bf x}_i$ are no more independent.
The one-- particle part of the Hamiltonian is independent of ${\bf X}$
and still exact. ${\bf X}=0$ only means making an approximation
to the e-- e-- interaction term in $H$.\\
It might be interesting to note that transformation (\ref{trafo}),
if applied to an Hamiltonian with an 
external Coulombic potential $\frac{Z}{r_i}$ (instead of the 
oscillator potential ${1\over 2}\omega_o^2 r_i^2$),
transforms in zero order in $\bf X$ the e-e-interaction term and
the potential energy in the external Coulombic potential into each other,
i.e. leaves the Hamiltonian in lowest order virtually invariant.

\newpage

\section{Exact solution in zero order in the center of mass coordinate}
\subsection{Pair Equation}

After expanding the kinetic energy, the Hamiltonian in zero order in
${\bf X}$  reads
\begin{equation}
H^{(0)}=\sum_{i=1}^3 h_i\;+\;H_{spin}
\end{equation}
with an effective pair Hamiltonian
\begin{equation}
h_i=-{1\over 2}{\bf \nabla}_i^2
+{1\over 2} \; \tilde\omega^2 \; x_i^2 + {1\over 2} \;\omega_c \; l_i
+\frac{1}{\sqrt{3}} \; {\beta \over |{\bf x}_i|} 
\label{h-i}
\end{equation}
where $\tilde\omega=\sqrt{\omega_0^2+(\frac{1}{2} \omega_c)^2}$,
$\omega_c=\frac{B}{c}$ is the cyclotron frequency, and $l_i$
is the orbital angular momentum operator.
This gives rise to the definition of an effective pair equation
\begin{equation}
h_i\; \phi_{q_i}({\bf x}_i) = \epsilon_{q_i} \; \phi_{q_i}({\bf x}_i)
\label{pair-eq}
\end{equation}
with the normalization condition
\begin{equation}
\int d^2 {\bf x}_i \; |\phi_{q_i}({\bf x}_i)|^2 =1.
\end{equation}
The subscript $q_i$ at eigenvalues and eigenfunctions 
comprises all quantum numbers
for the $i^{th}$ pair.
In polar coordinates $(x,\alpha)$, the pair equation (\ref{pair-eq}) 
is satisfied by the ansatz
\begin{equation}
\phi={e^{im\alpha}\over \sqrt{2\pi}} \; {u(x)\over
x^{1/2}}~~~~~;~~~~~m=0,\pm 1,\pm 2,\ldots
\label{ansatz}
\end{equation}
$u(x)$ must satisfy the {\em  radial pair equation}
\begin{equation}
\biggl\{-{1\over 2}~{d^2\over dx^2}+{1\over 2}\biggl(m^2-{1\over 4}\biggr)
{1\over x^2}+{1\over 2}\tilde\omega^2 x^2
+{\beta \over \sqrt{3} \; x}\biggl\}u(x)=\tilde\epsilon \; u(x)
\label{rad-eq}
\end{equation}
where $\tilde\epsilon=\epsilon-{1\over 2}m \omega_c$ 
and the solution is subject
to the normalization condition $\int\limits^\infty_o dx \;|u(x)|^2=1$.
In analogy to the two-- electron-- problem, we now
use for the radial part of the pair functions (\ref{ansatz})
the ansatz
\begin{equation}
u(x)=x^{|m|+\frac{1}{2}}\; p(x) \; e^{-\frac{1}{2}\; \tilde \omega \; x^2}
\end{equation}
where $p(x)$ is a polynomial, which is finite for solvable states.
In this way we obtain
for the pair function (\ref{ansatz}) in polar coordinates $(x,\alpha)$
\begin{equation}
\phi({\bf x})=\frac{{\bf x}^m }{\sqrt{2\pi}}  \; p(x)\;
e^{-\frac{1}{2}\; \tilde \omega \; x^2}
\label{pair-ansatz}
\end{equation}
where ${\bf x}^m$ is a shorthand  for $x^{|m|} e^{i m \alpha}$
(see also Sect.2).\\

Because of the decoupling in zero order,
the total eigenvalues and eigenfunctions of $H^{(0)}$ can be obtained from
\begin{equation}
E_{q_1,q_2,q_3}=\epsilon_{q_1}+\epsilon_{q_2}+\epsilon_{q_3}+E_{spin}
\label{E-tot}
\end{equation}
\begin{equation}
\Phi_{q_1,q_2,q_3}({\bf x}_1,{\bf x}_2,{\bf x}_3)=\phi_{q_1}({\bf x}_1)\cdot 
\phi_{q_2}({\bf x}_2)\cdot \phi_{q_3}({\bf x}_3)
\label{WF-tot}
\end{equation}

After inserting the transformation (\ref{diff}) into 
the spatial part of the total WF (\ref{WF-tot}),
we obtain in the original coordinates
\begin{eqnarray}
\Phi_{q_1,q_2,q_3}&=&
\phi_{q_1}\biggl({\bf R}-\frac{1}{\sqrt{3}}
({\bf r}_2-{\bf r}_3)\biggr)\cdot  \nonumber \\
& &\phi_{q_2}\biggl({\bf R}-\frac{1}{\sqrt{3}}
({\bf r}_3-{\bf r}_1)\biggr)\cdot  \nonumber \\
& &\phi_{q_3}\biggl({\bf R}-\frac{1}{\sqrt{3}}
({\bf r}_1-{\bf r}_2)\biggr)
\label{amended}
\end{eqnarray}
It is obvious that for any total WF the fields 
$\tilde \omega_{q_i}=\tilde \omega_q$ of the
three pairs have to agree.
If we 
insert (\ref{pair-ansatz}) into (\ref{amended}),
we end up with
\begin{eqnarray}
\Phi_{q_1,q_2,q_3} &=&
\biggl({\bf R}-\frac{1}{\sqrt{3}}
({\bf r}_2-{\bf r}_3)  \biggr)^{m_1} \;
p_{q_1}\biggl(\biggl|{\bf R}-\frac{1}{\sqrt{3}}
({\bf r}_2-{\bf r}_3) \biggr| \biggr)\cdot \; \nonumber \\
& &\biggl({\bf R}-\frac{1}{\sqrt{3}}
({\bf r}_3-{\bf r}_1)\biggr)^{m_2} \;
p_{q_2}\biggl(\biggl|{\bf R}-\frac{1}{\sqrt{3}}
({\bf r}_3-{\bf r}_1)\biggr) \cdot\; \nonumber \\
& &\biggl({\bf R}-\frac{1}{\sqrt{3}}
({\bf r}_1-{\bf r}_2)\biggr)^{m_3} \;
p_{q_3}\biggl(\biggl|{\bf R}-\frac{1}{\sqrt{3}}
({\bf r}_1-{\bf r}_2)\biggr|\biggr) \cdot  \nonumber \\
& & \mbox{exp}\biggl(-\frac{1}{2} \; \tilde \omega_q \sum_i {\bf r}_i^2 \biggr)
\label{amended-WF}
\end{eqnarray}
In order to simplify the exponential factor to the present form
we used the special property of our orthogonal transformation 
$\sum_i {\bf x}_i^2=\sum_i {\bf r}_i^2$.
This is still the most general result. 
It holds for ground and excited state and can be simplified in special cases.
For analytically solvable {\em finite-- field solutions}
the quantum numbers of all three pairs have to agree,
because the solvable fields $\tilde \omega_{q}$ for each pair depend on all
quantum numbers contained in  $q$. This implies in particular, that
all angular momenta $m_i$ in (\ref{amended-WF}) agree.
For {\em asymptotic solutions},  the solvable $\tilde \omega$
does not depend on
the angular momentum $m$ and the node number $k$
(or termination index $n=2k+1$) of the corresponding pair.
Therefore,  we can construct analytical solutions from pair states with
different quantum numbers, i.e. different $m_i$ and $k_i$.
For asymptodic ground state 
solutions, which belong to $n_i=1$, (\ref{amended-WF})
can be simplified by omitting the polynomials, because they are constants.\\

It should be mentioned that the eigenfunctions (\ref{WF-tot}) of $H^{(0)}$
form a complete set
and therefore they can be used as a basis for a numerical solution of the full
Schr\"odinger equation. Their advantage compared with basis functions
constructed from one-- particle states \cite{Girvin} 
\cite{Maksym} \cite{Hawrylak} \cite{Laughlin} is that they contain an e-- e--
correlation cusp. As well known in Quantum Chemistry,
the lack of this cusp in the basis functions gives rise to poor convergence
in CI expansions.  In the
present paper, however, we consider perturbation theory for improving the zero
order result beyond the strong correlation limit ${\bf R}=0$.

\subsection{Pauli Principle}
The WF (\ref{amended-WF}) does not yet 
fulfill the antisymmetry requirement.
Taking care of this is particularly important for finding the exclusion
principles for our type of WF. The question is: for what combination of quantum
numbers and parameter values does the antisymmetrized WF vanish 
by permutation symmetry? 
In order to keep formula length under control, we introduce some shorthands
and conventions. Firstly, the exponential factor in (\ref{amended-WF}) 
is fully
symmetric with respect to permutations and therefore it can be
 simply omitted in the
antisymmetrization procedure. Further, for the polynomial prefactor in 
(\ref{pair-ansatz}) we introduce a single symbol
\begin{equation}
t({\bf x})={\bf x}^m \; p(x)
\label{def-t}
\end{equation}
Then the spatial total WF can be written as
\begin{eqnarray}
\Phi_{q_1,q_2,q_3}({\bf r}_1,{\bf r}_2,{\bf r}_3)&=&
T_{q_1,q_2,q_3}({\bf r}_1,{\bf r}_2,{\bf r}_3) \cdot
\mbox{exp}\biggl(-\frac{1}{2} \; \tilde \omega_q \sum_i {\bf r}_i^2 \biggr)
\nonumber \\
T_{q_1,q_2,q_3}({\bf r}_1,{\bf r}_2,{\bf r}_3)&=&
t_{q_1}\biggl({\bf R}-\frac{1}{\sqrt{3}}({\bf r}_2-{\bf r}_3)\biggr)\cdot 
\nonumber \\
& &t_{q_2}\biggl({\bf R}-\frac{1}{\sqrt{3}}({\bf r}_3-{\bf r}_1)\biggr)\cdot 
\nonumber\\
& &t_{q_3}\biggl({\bf R}-\frac{1}{\sqrt{3}}({\bf r}_1-{\bf r}_2)\biggr) 
\label{T-WF}
\end{eqnarray}
Now, only the polynomial prefactor $T({\bf r}_1,{\bf r}_2,{\bf r}_3)$
has to be  antisymmetrized. \\

The implementation of the antisymmetrization procedure and the
antisymmetrized WFs can be found in the Appendix. The most
important qualitative result is that a simple 
{\em exclusion principle} exists only for ${\bf R}={\bf 0}$.
It states that the WF  vanishes by permutation symmetry:\\
in {\bf quartet} states $S=\frac{3}{2}$: \\
{\em if at least two pair functions agree and
if the total orbital angular momentum $M_L$ is even.}\\
in {\bf doublet} states $S=\frac{1}{2}$:\\
 {\em if all three pair functions  agree.}\\

Because for analytical {\em finite-- field-- solutions }
all quantum numbers of the 3 pairs have to agree, 
it follows from  the Pauli principle (see above), that only
quartet states (with parallel spins) can be given analytically.
If $m$ is the angular
momentum of the pair functions in (\ref{amended-WF}) then the total
orbital angular momentum in the solvable states is $M_L=3 m$.
In zero order on $\bf R$, Pauli principle demands that
$m=odd$. For finite $\bf R$ there is no restriction
from the Pauli principle and $m=integer$.
As will be shown in Section 6, it turns out that these  solvable 
states are  states where {\em kinks} in the curve $E$ versus $M_L$ occur
and which bear the so called {\em magic} angular momenta (see Fig. 2).
In short, only particular states can be solved analytically,
but these states are the most interesting ones.\\
For {\em asymptodic solutions} (with $\beta^2/ \tilde \omega=0$) 
all quantum numbers can be different,
although the pair functions belong to the same external fields. This means that
we can form total WFs from different pair functions and Pauli principle
applies as given above. \\
\begin{figure}[!bth]
\vspace{-2cm}
\begin{center}
{\psfig{figure=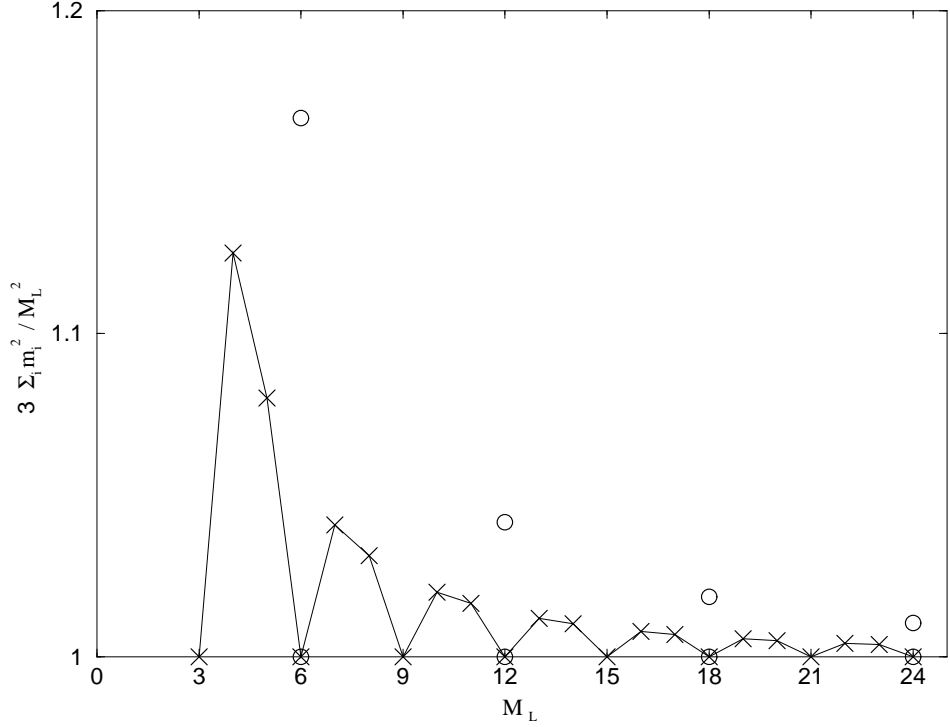,angle=-90,width=15.cm,bbllx=15pt,bblly=45pt,bburx=580pt,bbury=750pt}}
\caption[ ]{
The second term of the total energy in second order Taylor approximation
(\ref{E-Taylor}) divided by $c_2\; M_L^2 \;/3$
as a function of $M_L$ for the ground states (crosses).
Crossed  circles denote states which are forbidden by Pauli principle
for ${\bf R}=0$
(Laughlin approximation), but allowed
for finite ${\bf R}$.
The states denoted by open circles
are the ground states for ${\bf R}=0$ if the states underneath
are forbidden.
The region $M_L <3$ is omitted because it lies partly outside the scale and
because it needs special consideration.
}
\label{fig2}
\end{center}
\end{figure}

\subsection{Exact analytical solutions of the pair equation}
If we introduce in the radial equation
for the relative coordinate in the two-- electron-- case (\ref{rad-SGl})
the same parameters as used for 3 particles (i.e.
$\tilde \omega$ and $\tilde \epsilon$), we obtain
\begin{equation}
\biggl\{-{1\over 2}~{d^2\over dr^2}+{1\over 2}\biggl(m^2-{1\over 4}\biggr)
{1\over r^2}+{1\over 2} \biggl( \frac{\tilde\omega^{(N=2)}}{2} \biggr)^2 r^2
+{\beta\over 2 \; r}\biggl\}u^{(N=2)}(x)=
\biggl( \frac{\tilde\epsilon^{(N=2)}}{2} \biggr) \; u^{(N=2)}(x)
\label{rad-eq2}
\end{equation}
For avoiding confusion, the parameters and the WF 
in the two-- particle-- problem have been
given the extra superscript $N=2$.
Comparison with the radial pair equation (\ref{rad-eq}) shows us that
we obtain solutions of (\ref{rad-eq}) from solutions of (\ref{rad-eq2})
by simple rescaling.\\
The 'spectrum' of soluble $\tilde \omega$ follows from
\begin{equation}
\tilde\omega=\frac{2}{3}\;\tilde\omega^{(N=2)}
\label{sol-om}
\end{equation}
and the corresponding eigenvalues and eigenfunctions from
\begin{equation}
\tilde\epsilon=\frac{2}{3}\;\tilde\epsilon^{(N=2)}
\label{sol-eps}
\end{equation}
\begin{equation}
u(x)=\sqrt{\frac{2}{\sqrt{3}}} \;\;
 u^{(N=2)}\biggl(r=\frac{2}{\sqrt{3}} \; x\biggr)
\label{sol-wf}
\end{equation}
The prefactor in (\ref{sol-wf}) has been chosen to retain normalization
of the radial pair function, if $u^{(N=2)}(r)$ is normalized.
In any solution, the quotient $\frac{\tilde\epsilon}{ \tilde\omega}$
is equal in both problems and given by (see \cite{Taut2e})
\begin{equation}
\frac{\tilde\epsilon}{\tilde\omega}=
\frac{\tilde\epsilon^{(N=2)}}{\tilde\omega^{(N=2)}}=|m|+n
\label{sol-quotient}
\end{equation}
where $m$ is the angular momentum and $(n-1)$ the highest power
in the polynomial $p(x)$.\\
In this way we obtain from (\ref{om-N=2-n=1} -- \ref{u-N=2-n=3}) and
(\ref{sol-om} -- \ref{sol-wf})
the simplest exact solutions of  
(\ref{rad-eq}) as follows:\\
 For $n=1$ there is only the 
asymptotic solution
\begin{eqnarray}
\frac{\beta^2}{\tilde \omega}&=&0\\
p(x)&=&1
\end{eqnarray}
For $n=2$ there is one finite-- field-- solution:
\begin{eqnarray}
\frac{\beta^2}{\tilde\omega}&=&\frac{3}{2}\; (2\; |m|+1) \label{om-n=2}\\
p(x) &=&  1+\frac{\frac{2}{\sqrt{3}}\; \beta  x} {(2\; |m|+1)} 
\label{u-n=2}
\end{eqnarray}
Both former solutions are ground states.
For $n=3$ there is one asymptotic solution, which is a first excited state,
\begin{eqnarray}
\frac{\beta^2}{\tilde \omega}&=&0\\
p(x)&=&1-\frac{\tilde \omega\; x^2}{(|m|+1)}
\end{eqnarray}
and one at finite fields, which is a ground state,
\begin{eqnarray}
\frac{\beta^2}{\tilde\omega}&=&3\; (4\; |m|+3) 
\label{om-n=3}\\
p(x) &=&  1+\frac{\frac{2}{\sqrt{3}}\;\beta x} {(2\; |m|+1)} 
          +\frac{\biggl(\frac{2}{\sqrt{3}}\;\beta x \biggr)^2} 
                {2 (2\; |m|+1)(4\;|m|+3)} 
\label{u-n=3}
\end{eqnarray}
The corresponding energies follow from (\ref{sol-quotient}), and
we put the normalization constant 
$a_0=1$  without loss
of generality. The pattern of solvable states agrees qualitatively
with Fig.1. Only the abscissa-- values of solvable states are shifted.
Those, who are not yet convinced in the correctness of these solutions
 are recommended to check them
 by insertion into   (\ref{rad-eq}).

\newpage

\section{Comparison of our analytical solutions with Quantum Hall States}
In this section we confine ourselves to ground states and consider 
two cases  separately.
This will provide two different generalizations of the Laughlin WFs.
We want to emphasize, however, that it would be  possible 
(but not convenient) to include both cases
in one formula.
Secondly, we start with considering the case $\bf R=0$ (zero order result)
and add some words on the general case afterwards. 
Generally, the generalized formulae in this section comprise our
analytical results 
for 2 particles and 3 particles. However, they could {\em ad hoc} be applied
to any particle number and considered as   trial functions.\\

The {\em first case} comprises all solutions for three electrons with 
equal  pair-- angular-- momenta in (\ref{amended-WF})
and the two-- electron result (\ref{tot-WF-2e}). 
In other words, it includes all finite-- field solutions and and 
the asymptodic solutions with equal  pair-- angular-- momenta.
It can be written in the following unified form
\begin{equation}
\Phi = \prod _{i<k} ({\bf r}_i-{\bf r}_k)^m\;
p_{n,m}(|{\bf r}_i-{\bf r}_k|)\;\;
\mbox{exp}\biggl(-\frac{1}{2} \; \tilde \omega_{n,m} \sum_l {\bf r}_l^2 \biggr)
\label{amended-general}
\end{equation}
It should be remembered that,  apart from the case $n=1$, 
the soluble field values $\tilde \omega_{n,m}$
and the polynomials $p_{n,m}(x)$
depend on the particle number.\\
Using complex coordinates as defined in Section 2, (\ref{amended-general})
can be reformulated as
\begin{eqnarray}
\Phi &=& \prod _{i<k} (\bar z_i-\bar z_k)^{\tilde m} \;
p_{n,m}(|\bar z_i-\bar z_k|) \; \chi_1(\tilde\omega_{n,m}) 
~~~~~\mbox{for}~~~m\le 0 
\nonumber                                 \\
     &=& \mbox{complex conj.}~~~~~  \mbox{for}~~~ m\ge 0 
\label{amended-chi1}
\end{eqnarray}
where $\chi_1(\tilde\omega_{n,m})$ is a Slater determinant 
of LLL states for  the effective frequency
$\tilde\omega_{n,m}$ and $\tilde m =|m|-1$. For $m\le 0$ our WF can 
also be rewritten using the Laughlin WF \cite{Laughlin}
\begin{equation}
\Phi = \prod _{i<k} 
p_{n,m}(|\bar z_i-\bar z_k|) \; 
\Phi^{Laughlin}_{\nu=\frac{1}{\bar m}}(\tilde\omega_{n,m})
\label{amended-L}
\end{equation}
where the fields $\tilde\omega_{n,m}$ occur in the exponential factors
of the Laughlin function instead of the infinite field used
in the original Laughlin function.\\
Our solutions (\ref{amended-general}) and the 
equivalent forms have the following properties:
$\Phi$ fulfills the Pauli principle, if $m=odd$ (and $\tilde m=even$).
It is an eigenfunction of the total angular momentum operator with eigenvalue
$M_L=m \frac{N(N-1)}{2}$, where $N$ is the electron number.
Apart from the asymptodic case $n=1$, it has components in higher LLs due to the
$p(x)$--factors.
In the case $n=1$ (where $p(x)=const$ and $\frac{1}{\tilde \omega_1}=0$)
 and $m \le 0$, our WFs agree with the
Laughlin states.
\begin{equation}
\Phi^{Laughlin}_{\nu=\frac{1}{\bar m}}(\tilde\omega_1)=
\prod _{i<k} (\bar z_i-\bar z_k)^{\bar m} \;
\mbox{exp}\biggl(-\frac{1}{2} \; \tilde \omega_1 \sum_l |{\bar z}_l|^2 \biggr)
= \prod _{i<k} (\bar z_i-\bar z_k)^{\tilde m} \; \; \chi_1(\tilde\omega_1)
\label{Laughlin}
\end{equation}
For comparison, we also quote the Jain ansatz \cite{Jain-FQHE}
\begin{equation}
\Phi_{\nu}^{Jain}= {\cal P}_{LLL} \; 
\prod _{i<k} (\bar z_i-\bar z_k)^{\tilde m} \; 
\chi_{\nu^*}(\tilde\omega_1)
\label{Jain}
\end{equation}
where $\frac{1}{\nu}=\pm \frac{1}{\nu^*}+\tilde m$,
${\cal P}_{LLL}$  is a projection operator 
onto the LLL and the determinant
$\chi_{\nu^*}(\tilde\omega_1)$ is for $\nu^*$ full LLs and taken at 
the asymptotically infinite frequency $\tilde\omega_1$. 
Apart from the special cases discussed above, (\ref{amended-chi1})
and (\ref{Jain}) do  not seem to be fully equivalent. At least, 
both treatments contain the Laughlin WF as a special case, and
there is the vague similarity that both  have to do with higher LL
components.\\
The next property will be discussed using the formulae for $N=3$, but it would
also apply to the corresponding generalized trial functions.
If we go beyond the ${\bf R}=0$ approximation, i.e. if we calculate the
 {\em form} of the pair functions in the transformed 
space ${\bf x}_i$ in zero order
in ${\bf X}={\bf R}$, but use the full back-- transformation 
to the original coordinates ${\bf r}_i$ involving
a {\em finite} ${\bf R}$, then  the solvable eigenfunctions  are those in
(\ref{amended-WF}) with $m_1=m_2=m_3=m$. Firstly, we have to remind that 
this function has to be antisymmetrized as discussed in the Appendix,
because it is not automatically antisymmetric for $m=odd$ as
in the case ${\bf R}=0$.
This provides a complicated expression. Due to this antisymmetrization
it has simple zeros wherever two coordinates agree. However, there are no
$m$-- fold zeros as in the Laughlin WF, because the factors
$\biggl({\bf R}-\frac{1}{\sqrt{3}} ({\bf r}_i-{\bf r}_k)  \biggr)$
do not vanish if two coordinates agree. This holds for all solutions
(finite and infinite field solutions). This is a feature which agrees 
qualitatively with
the Jain functions.\\

The {\em second case} differs from the first case only for $N>2$.
For the asymptodic solutions in (\ref{amended-WF}), different $m_i$ are allowed,
and for the ground state the polynomials can be omitted because they are 
constants. After introducing  a new indexing for the angular momenta,
which is more appropriate for the $N$--particle system, 
we obtain from (\ref{amended-WF})
\begin{eqnarray}
\Phi &=& \prod _{i<k} ({\bf r}_i-{\bf r}_k)^{m_{ik}}\;\;
\mbox{exp}\biggl(-\frac{1}{2} \; \tilde \omega_1 \sum_l {\bf r}_l^2 \biggr)
\label{amended-different-m}\\
&=&\prod _{i<k} (\bar z_i-\bar z_k)^{{\bar m}_{ik}} \;
\mbox{exp}\biggl(-\frac{1}{2} \; \tilde \omega_1 \sum_l |{\bar z}_l|^2 \biggr)
\end{eqnarray}
The second equation holds for $m_{ik} \le 0$.
For general $m_{ik}$, this function has to be antisymmetrized. For 3
particles, the antisymmetrized result is discussed in Section 4.2 and 
given in the Appendix. In this special case Pauli principle
demands that $M_L=odd$.
$\Phi$ in (\ref{amended-different-m}) is an eigenfunction of the total
orbital angular momentum with eigenvalue $M_L=\sum_{i<k} m_{ik}$,
and lies completely within the LLL.
It is apparent that the special case of equal $m_{ik}=m=odd$ agrees
with the Laughlin states.  The changes produced by the extension 
of the considerations to finite $\bf R$ are analogous to the first case.\\

A more detailed investigation of
the applicability of the two generalizations  (\ref{amended-general}) 
and (\ref{amended-different-m}) to $N$--particle systems and in particular
Quantum Hall states
will be the aim of a separate work. \\

\newpage

\section{Approximate analytical solution of the pair equation
and magic angular momenta for three electrons}
In order to avoid misunderstandings, it should be told at the very beginning,
 why we are looking for approximate solutions, if there are exact ones.
The answer is that the exact solutions
exist only for certain fields and states,
whereas the solutions of this subsection are for all parameters and all states.
Although the exactly soluble states are the most interesting ones, the rest is
necessary to prove certain cusp properties of the exactly solvable ones.
From here on we consider only the case  $\beta=1$ and finite fields.\\
As shown in \cite{Taut2e}, sect.3.2, the effective potential in
the radial pair equation (\ref{rad-eq}) can be expanded around its minimum
into a Taylor series to second order.
This cannot be accomplished fully analytically because the minimum position
results from the zeros of a forth order polynomial equation.
It is possible, however, to establish the approximate effective potential
to order $r_0^{-1}$, where $r_0^{-1}=(\frac{9}{8} \tilde \omega)^{1/3}$
is a small parameter in the strong correlation limit.
The eigenvalues of the resulting oscillator equation
can then be given analytically with the result
\begin{equation}
\epsilon= c_0 + c_1 \; m + c_2 \; m^2 + O(r_0^{-2})\\
\end{equation}
where
\begin{eqnarray}
c_0&=&\biggl[ 1-\frac{1}{4\;\; 3^{1/3}} \tilde\omega^{2/3}\biggr]\;
\biggl[ \frac{1}{2} \; (3\; \tilde \omega)^{2/3} + \sqrt{3}\; \tilde \omega
         \;(k+\frac{1}{2}) \biggr] \\
c_1&=&\frac{1}{2} \; \omega_c \\
c_2&=&3^{-1/3}\; \tilde\omega^{2/3}\;
\biggl[ \frac{1}{2} \; (3\; \tilde \omega)^{2/3} + \sqrt{3}\; \tilde \omega
         \;(k+\frac{1}{2}) \biggr]
\end{eqnarray}
where $k$ is the node number.
This provides   a total energy of
\begin{equation}
E=3\;c_0 + c_1\; M_L +c_2\;(m_1^2+m_2^2+m_3^2)
\label{E-Taylor}
\end{equation}
where $M_L=m_1+m_2+m_3$.
 It is clear that the
Taylor expansion is the better the more symmetric the effective potential is.
 Therefore it gets poorer with increasing $m$ and increasing $\tilde \omega$.
Nevertheless, this formula gives a
qualitative understanding of the magic angular momenta. If we are interested
in the ground state {\em for a given} $M_L$, it is clear 
from (\ref{E-Taylor})
that it is formed by
that set of $m_i$ for which
the sum of the squares of the $m_i$ is minimal for a given sum of the $m_i$.
This demand is met if the $m_i$ are 'as equal as possible'.
As a example,
for $M_L=2$ there are two sets which provide the same $M_L$, namely
$(002)$ and $(011)$, with the latter forming the ground state.
It is also clear that the total momenta of the form $M_L=3 m$ 
(multiple of three) play a special role because all three $m_i$ can be
equal in this case, but not for the other $M_L$.
Fig. 1 shows the third (and only discontinuous) term 
of (\ref{E-Taylor}) as a function of $M_L$.
It is obvious that this curve has kinks whenever it is possible that
all three $m_i$ are equal, i.e. for $ M_L= 3 m$ with $m=0,1,2, \cdots$.
On the other hand, we learned, that equal $m_i$ is a prerequisite for
analytical solutions. Thus we conclude that {\em the solvable states are
always cusp states}.\\
Observe, that in the limit ${\bf R}=0$ the  Pauli principle 
demands that even $m$
are forbidden (see Sect. 4.2). 
We also want to remind (see Sect. 5) that we obtain the Laughlin WF
in the ${\bf R}=0$ limit.
If we go beyond this approximation, the total WFs (\ref{amended}) 
can be antisymmetrized for {\em any} values of the $m_i$. 
These facts elucidate the origin for the well known problem that
exact diagonalization procedures provide cusps also for those $M_L$, 
which correspond to even denominator Laughlin states 
\cite{Girvin}.
This shows that the Laughlin ansatz (if applied to finite systems)
has an additional symmetry (produced by putting ${\bf R}=0$), which
is {\em not} present in exact solutions.
These conclusions are not in contradiction with the symmetry considerations
in \cite{Ruan} and \cite{Seki}. The latter papers find formula for
the cusp (or magic) angular momenta, provided, the states under
investigation are not
forbidden by Pauli's principle. They do not have (and need) any explicit 
expression for the wave function. By the way, their general notion on the 
eigenstates is consistent with our small--$\bf R$ expansion.
The harmonic approximation used in this section is also related to the
harmonic approach used in \cite{Maksym}, but not fully equivalent. However,
our analytic approximations derived in  Sect.s 3 and 4 is not harmonic.

\newpage

\section{First order perturbation theory and accuracy of the strong
correlation expansion}

Now we are going to calculate the contributions of the dipole and quadrupole
term (\ref{dipole}) and (\ref{quadrupole})  of the e-e-interaction
to the total energy in first order
\footnote{By first order we mean first order in
all multi-pole corrections, but not
first order in $\bf R$}
perturbation theory, i.e.
\begin{equation}
\Delta E^{(l)}=  < \Phi| V_{ee}^{(l)} |\Phi>
\label{E-first-order}
\end{equation}
where the zero order wave function is generally given by (\ref{tot-WF-sym}).
For {\em quartet states}, where the total WF is just a product of spatial and
spin part and for calculating matrix elements, the total WF in
(\ref{E-first-order}) can be replaced by the unsymmetrized spatial part.
(It is simpler, however, to do the integrations in the
transformed ${\bf x}_i$--coordinates rather than in the ${\bf r}_i$).
The further calculation is straight forward and provides
\begin{eqnarray}
\Delta E ^{(1)}&=& \frac{1}{3\sqrt{3}} \; \sum_{k=1}^3 M_{m_k}^{(-1)} \\
\Delta E ^{(2)}&=& \frac{1}{9\sqrt{3}} \;\biggl[ \sum_{k=1}^3 M_{m_k}^{(-1)}-
\frac{1}{2}\; \sum_{k=1}^3 \sum_{l(\neq k)=1}^3 
M_{m_k}^{(-3)}\;M_{m_k}^{(2)} \biggr]
\end{eqnarray}
where we defined moments
\begin{equation}
M_m^{(k)}=\int_0^{\infty} dx \; x^k \; [u_m(x)]^2
\label{moment}
\end{equation}
with $u_m(x)$ being the radial part defined in (\ref{ansatz}) and given
explicitly in special cases using (\ref{u-n=2}) and (\ref{u-n=3}).
If all three angular momenta agree:  \mbox{$m_1=m_2=m_3\equiv m$},
the result simplifies to
\begin{eqnarray}
\Delta E ^{(1)}&=& \frac{1}{\sqrt{3}} \;  M_{m}^{(-1)}
\label{E1}\\
\Delta E ^{(2)}&=& \frac{1}{3\sqrt{3}} \;\biggl[  M_{m}^{(-1)}-
        M_{m}^{(-3)}\;M_{m}^{(2)} \biggr]
\label{E2}
\end{eqnarray}
It should be noted that the moment $M_{m}^{(-3)}$, appearing in the second
order contribution, diverges for $m=0$. This is because for $x \rightarrow 0$
the radial pair function  goes as
$u_m(x) \rightarrow  x^{(|m|+\frac{1}{2})}$  and
thus the integrand in (\ref{moment}) behaves
like $x^{-2}$ for small $x$. That is why the results for $m=0$ are missing in
Table 2.\\

For a test of the accuracy of our small--$\bf R$ expansion
we use analytically solvable states only, i.e. quartet states
with $M_L=3 m$. We do this 
for magnetic field only (i.e. $\omega_0=0$), 
because the confinement can be included afterwards by
a simple rescaling of the parameters.
For $\omega_0=0$ it follows from (\ref{sol-quotient}) and the definitions of
$\tilde \omega$ and $\tilde \epsilon$ that the zero order 
(in $\bf R$) energy per electron
reads
\begin {equation}
\frac {E^{(0)}}{3 \omega_c}=\frac{(m + |m|)}{2}+\frac{n}{2}
\label{E-exact}
\end{equation}
and, for comparison, the trivial result without e-e-interaction is
\cite{one-e-dot}
\begin {equation}
\frac {E^{(non-int)}}{3 \omega_c}=\frac{(m + |m|)}{2}+k+\frac{1}{2}
\end{equation}
The formulae for the corrections in  first and second order $\Delta E ^{(1)}$
and $\Delta E ^{(2)}$, respectively,
are given in (\ref{E1}) and (\ref{E2}). 
Table 1 and 2 show the results for fixed $m$ and
varying $n$ and fixed $n$ and different $m$, respectively.
$E^{(Taylor)}$ is the result using the Taylor expansion 
of the effective potential in the radial Schr\"odinger equation as
 described in Sect. 6.
Its agreement with the exact (analytical) solution of the radial 
Schr\"odinger equation $E^{(0)}$ gives an account of the accuracy of the Taylor
expansion.\\

\begin{table}[]
\caption[]{
Total energy per electron in units of $\omega_c$
for analytically solvable quartet states
with total orbital angular momentum $M_L=-3$ ($m=-1$)
for $\omega_0=0$ (magnetic
field only). $E^{(non-int)}$ is the energy without
electron-- electron interaction,
$E^{(Taylor)}$ is the total energy where the
effective potential in the radial pair equation is treated
in second order Taylor expansion (\ref{E-Taylor}), and
$E^{(0)}$ is the exact result in zero order in $\bf R$
as given in (\ref{E-exact}).
 $\Delta E^{(1,2)}$ are the
dipole and quadrupole contributions
given in (\ref{E1}) and (\ref{E2}), and $E^{(2)}$ is
the sum of the three former contributions.
The Zemann energy is omitted.
\vspace{.5cm}}

\renewcommand{\baselinestretch}{1.5}
\begin{tabular}{|r|r|c|c|l|r|r|r|}\hline
n& $\omega_c$ & $\frac {E^{(non-int)}}{3 \omega_c}$ &
$\frac {E^{(Taylor)}}{3 \omega_c}$ &
$\frac {E^{(0)}}{3 \omega_c}$ &
$\frac {\Delta E^{(1)}}{3 \omega_c}$ &
$\frac {\Delta E^{(2)}}{3 \omega_c}$ &
$\frac {E^{(2)}}{3 \omega_c}$  \\ \hline
2 & $\frac{4}{9}$= 444.444 E-3 & 0.5 & 1.03796
& 1 & 0.154332 & --0.101713 & 1.052619  \\
3 & $\frac{2}{21}$=95.2381 E-3 & 0.5& 1.49531
 & 1.5 & 0.290373 & --0.126757 & 1.663616\\
5 & 18.1896 E-3              & 0.5 & 2.49055
& 2.5 & & & \\
10& 2.20940 E-3               & 0.5 & 4.99381 & 5   & & & \\
15& 0.655360 E-3              & 0.5 & 7.49561 & 7.5 & & & \\ \hline
\end{tabular}
\end{table}
\begin{table}[]
\caption[]{
Total energy per electron in units of $\omega_c$
for analytically solvable quartet states
for $\omega_0=0$ (magnetic field only) as a function of $m$ with
$M_L=3 m$. ({\em Only odd $m$ are compatible with the Pauli principle}.)
All solutions belong to $n=2$.
The meaning of the other column heads is as in Table 1.
\vspace{.5cm}}

\renewcommand{\baselinestretch}{1.5}
\begin{tabular}{|r|r|c|c|r|r|l|r|r|}\hline
m& $\omega_c$ & $\frac {E^{(non-int)}}{3 \omega_c}$ &
$\frac {E^{(Taylor)}}{3 \omega_c}$ &
$\frac {E^{(0)}}{3 \omega_c}$ &
$\frac {\Delta E^{(1)}}{3 \omega_c}$ &
$\frac {\Delta E^{(2)}}{3 \omega_c}$ &
$\frac {E^{(2)}}{3 \omega_c}$ \\ \hline
10 & $\frac{4}{63}$=0.063 & 10.5 & 21.4662
& 11 & 0.164717 & --0.00834761 &  11.15637 \\
...&&&&&&\\
3 &  $\frac{4}{21}$=0.190 & 3.5 & 5.06036
 & 4 & 0.161018 & --0.0283146 & 4.132703 \\
2 &  $\frac{4}{15}$=0.266 & 2.5 & 3.43552
 & 3 & 0.158918 & --0.0435331 & 3.115385 \\
1 &  $\frac{4}{9}$=0.444 &  1.5 & 2.03796
& 2 & 0.154332 & --0.101713 & 2.052619 \\
\hline
0 & $\frac{4}{3}$=1.333 & 0.5 & 0.892263
& 1 & 0.136400 &&\\
\hline
--1 &  $\frac{4}{9}$=0.444 &  0.5 &1.03796
 & 1 & 0.154332 & --0.101713 & 1.052619 \\
--2 &  $\frac{4}{15}$=0.266 & 0.5 & 1.43552
 & 1 & 0.158918 & --0.0435331 & 1.115385 \\
--3 &  $\frac{4}{21}$=0.190 & 0.5 & 2.06036
 & 1 & 0.161018 & --0.0283146 & 1.132703 \\
...&&&&&&\\
--10 & $\frac{4}{63}$=0.063 & 0.5 & 11.4662
& 1 & 0.164717 & --0.00834761 &  1.156369 \\
\hline
\end{tabular}
\end{table}

We conclude the following.\\
i) Comparison of $E^{(non-int)}$ and $E^{(0)}$ shows the contribution of the
Coulomb interaction energy to the total energy. As to be expected, 
it grows with
decreasing $\omega_c$ and is tremendous for small $\omega_c$ (e.g.
ten times larger than the kinetic energy
for $n=10$ i.e. $\omega_c \approx 10^{-3}$).\\
ii) For small $|m|$, the Taylor approximation
provides a good tool for solving the radial Schr\"odinger equation. 
For large $|m|$ the effective potential becomes so unsymmetric 
with respect to its minimum that its result goes fatally wrong. \\
iii) The  analytical zero order result $E^{(0)}$, which is the main 
achievement of this paper, compares pretty well with the most precise
result $E^{(2)}$. This is  mainly due to the 
cancellation of $\Delta E ^{(1)}$ and
$\Delta E ^{(2)}$  . The maximum error is of the order of 10 \%.

\section{Summary and discussion}
We found that the two--dimensional 
Schr\"odinger equation for 3 electrons in an homogeneous
magnetic field (perpendicular to the plane) and a parabolic 
scalar confinement potential has exact analytical solutions in the
strong correlation limit, where the expectation value of the 
center of mass vector is small compared with the 
average distance between the electrons.
These analytical solutions exist only
for certain discrete values of the external fields
$\tilde \omega$. 
For finite external fields, 
analytical solvability demands that
all three pair-- angular-- momenta agree what leads 
to total angular momenta $M_L=3 m$ with $m=integer$.
In zero order in $\bf R$, Pauli principle allows 
equal pair-- angular-- momenta 
only for parallel spins and $m=odd$. The analytically solvable states are always
cusp states, i.e. states where $E(M_L)$ has a cusp.\\
Further, these special 
analytical solutions for 3 particles and the exact analytical solutions for 
2 particles could be written in 
a unified form, which contains only products over coordinate combinations
$\prod _{i<k}$  and sums $\sum_i$.
Conveniently, instead of using one formula we consider
two cases   (\ref{amended-general}) and
(\ref{amended-different-m}). 
These formulae, when {\em ad hoc} generalized to  N coordinates,
can be discussed as ansatzes for the wave function of the N--particle system.
These ansatzes fulfill the following demands: they are exact for two particles
and for 3 particles in
the limit of small $\bf R$,
and they are eigenfunction of the total
orbital angular momentum. The Laughlin functions are special case,
or in other words, both formulae provide two different generalizations of the
Laughlin functions.\\
Until now we know mainly that our WFs are analytically solvable states
(within the approximations discussed above). It is also clear that
states for an infinitesimally  varied $\tilde \omega$ look quite different,
i.e. the finite polynomials have to be replaced by polynomials
with an infinite number of terms. For N=2 and 3 these 
'neighboring' states are even explicitly known. It is not yet shown,
however, if these special states have generally  something to do with
the Quantum Hall states (which show  similar singular features), 
or if any physical quantity has any special feature in these 
soluble states.
One encouraging fact is that the Laughlin states are special cases of
our states. 
In prospect, it is possibly a good idea to look for 
similar exact analytical solutions 
in a spherical geometry instead of the disk geometry used here, because 
it is easier then to attribute a filling factor to each eigensolutions.
If there is a connection between our states and Quantum Hall states,
this would imply some kind of 
inherent super-symmetry in the Quantum Hall states.\\
Now we want to compare our treatment of three electrons
with Laughlins \cite{Laughlin}. Both approaches are approximate.
Whereas he forms WF by antisymmetrization of one-- particle states of the
LLL, which is expected to be good for strong fields, 
we established an expansion,
which is good in the strong correlation limit and which contains
in general higher LL components. Consequently, we obtained a richer variety
of solutions comprising the Laughlin states as special cases. 

\newpage

\section{Appendix: Pauli Principle}

In order the obtain familiar looking formula, we rename $T \rightarrow \Phi$
and $t \rightarrow \phi$.
The question here is, how the properly symmetrized spatial part $\Phi$ 
has to be
supplemented by an appropriate spin part in order to obtain a wave function
which is eigenfunction of ${\bf S}^2$ and $S_z$ (with quantum numbers
$S$ and $M_S$) and which satisfies the Pauli principle.
${\bf S}=\sum_{i=1}^N {\bf s}_i$ is the total spin operator for all particles.
For more than two particles this is
a well established, but non-- trivial procedure.
The source of the difficulty is the fact that
the spin space can be degenerate, i.e. there is
more than one orthogonal spin eigenfunction ${\it X}_i,\; (i=1,...f)$ for
given $S$ and $M$ (see Table 3 for N=3).\\

\begin{table}[hb]
\caption[]{

Standard spin eigenfunctions for N=3.  $\alpha$ and $\beta$ are the
one--particle spin eigenfunctions for spin up and down, respectively.
The first factor of a pro\-duct of one--particle
functions carries the spin variable '1',
the second factor carries '2' etc.\\}

\renewcommand{\baselinestretch}{1.5}
\large
\begin{tabular}{|r|r|l|l|l|}\hline
S& $M_S$ & i  & ${\it X}_i$& permutation sym.  \\ \hline
$\frac{1}{2}$  & +$\frac{1}{2}$  & 1 & $\frac{1}{\sqrt{6}}$
[--(\a \b+\b \a)\a +2 \a \a \b ]& symmetric for (12) \\
               & --$\frac{1}{2}$  &   & $\frac{1}{\sqrt{6}}$
[+(\b \a+\a \b)\b --2 \b \b \a ]&                     \\ \hline
$\frac{1}{2}$  &+$\frac{1}{2}$   & 2 & $\frac{1}{\sqrt{2}}$
(\a \b -- \b \a) \a             &antisymmetric for (12)\\
               &--$\frac{1}{2}$   &   & $- \frac{1}{\sqrt{2}}$
(\b \a -- \a \b) \b             &                     \\ \hline
$\frac{3}{2}$  & +$\frac{3}{2}$  & 1 &
\a \a \a                             & symmetric for all {\cal P}\\
               & +$\frac{1}{2}$  &   & $\frac{1}{\sqrt{3}}$
[\a (\a \b+\b \a)+\b \a \a ]     & \\
               & --$\frac{1}{2}$  &   & $\frac{1}{\sqrt{3}}$
[\b (\b \a+\a \b)+\a \b \b ]     & \\
               & --$\frac{3}{2}$  &   &
\b \b \b                             & \\ \hline
\end{tabular}
\end{table}
\normalsize

Generally, the total WF $\Psi$, which fulfills our demands,
can be calculated from (see e.g. \cite{Pauncz})
\begin{equation}
\Psi_i(1,2,3)= {\cal A}_a \; {\it X}_i(1,2,3) \; \Phi(1,2,3)~~~~~~~~i=(1,... ,f)
\label{antisym}
\end{equation}
with the antisymmetrizer
\begin{equation}
{\cal A}_a=\frac{1}{\sqrt{N!}} \sum _{\cal P} (-1)^p \; {\cal P}
\end{equation}
and \cal $\cal P$ is a permutation operator.  $f$ is the dimension
of the degenerate spin space ($f=1$ and $2$ for
 $S=\frac{3}{2}$ and $\frac{1}{2}$,
respectively, for N=3) and $(-1)^p$ is the
parity of the permutation $\cal P$. The quantum numbers $S$ and $M_S$
 of the spin part ${\it X}_i$
as well as the quantum numbers of the spatial part $\Phi$
are not indicated. The arguments $(1,2,3)$ are
spin or spatial coordinates depending on the function in question.
Although being correct, (\ref{antisym}) does not reveal the inherent
permutation symmetry of the total WF. An equivalent symmetrized form is
(see e.g. \cite{Pauncz})
\begin{equation}
\Psi_i=\frac{1}{\sqrt{f}} \sum_{k=1}^f {\it X}_k \cdot \Phi^s_{ki}
\label{tot-WF-sym}
\end{equation}
where the symmetrized spatial function is defined as
\begin{equation}
\Phi^s_{ki}=\sqrt{\frac{f}{N!}}\sum _{\cal P} (-1)^p \; U_{ki}({\cal P})\;
{\cal P} \; \Phi
\end{equation}
and $U_{ki}({\cal P})$ is an irreducible representation matrix of the
permutation group for permutation $\cal P$ given in Table 4.
It is convenient to define  column vectors $\bf \Phi\mbox{$^s_i$}$
(with $i=1,2$) of
the matrix $\Phi^s_{ki}$. Then the $i^{th}$ vector
comprises all spatial information
about the $i^{th}$ of the  orthogonal states $\Psi_i$.\\

\begin{table}[]
\caption[]{
Irreducible representation matrices ${\bf U}({\cal P})$ for N=3. For
class  and cyclic permutation symbols see \cite{Pauncz}.\\}

\renewcommand{\baselinestretch}{1.5}
\large
\begin{tabular}{|c|c|c|}\hline
class& {\cal P} & {\bf U}({\cal P})   \\ \hline

[$1^3$] & $\varepsilon$ &
$ \left[ \begin{array}{cc} 1&0\\0&1 \end{array}\right] $ \\ \hline

[$2,1$] & (1,2) &
$ \left[ \begin{array}{cc} 1&0\\0&-1 \end{array}\right] $ \\ \cline{2-3}

        & (2,3) &
$ \frac{1}{2} \left[ \begin{array}{cc} -1& \sqrt{3} \\ \sqrt{3}&1 \end{array}
\right] $ \\
\cline{2-3}

        & (3,1) &
$ \frac{1}{2} \left[ \begin{array}{cc} -1&-\sqrt{3}  \\ -\sqrt{3}&1 \end{array}
\right]$\\ \hline

[$3$]  & (123)  &
$ \frac{1}{2} \left[ \begin{array}{cc} -1&\sqrt{3}\\-\sqrt{3}&-1 \end{array} 
\right] $ \\
\cline{2-3}

       & (132)  &
$ \frac{1}{2} \left[ \begin{array}{cc} -1&-\sqrt{3} \\ \sqrt{3} & -1 \end{array}
 \right] $ \\ \hline

\end{tabular}
\end{table}

{\bf Quartet $S=\frac{3}{2}$}\\
Because of $f=1$ and because the spin eigenfunctions are symmetric
against all permutations, the symmetrized spatial function is totally
antisymmetric and we have
\begin{equation}
\Psi={\it X}\; \Phi^s ;~~~~~~~~~~~~~~~\Phi^s={\cal A}_a \Phi
\end{equation}
which is reminiscent of the permutation
symmetry of the triplet state for $N=2$.
If we insert the  solution (\ref{T-WF}) for $\Phi$
we obtain a lengthy expression which does not reveal anything.
For ${\bf R}={\bf 0}$, however,
(\ref{def-t}) implies
$\phi_{m}(-{\bf x})=(-1)^m \; \phi_{m}({\bf x})$ or
in the short hand notation
\begin{equation}
\phi_{m}(i-k)=(-1)^m \; \phi_{m}(k-i)
\label{inversion}
\end{equation}
 with $m$ the orbital angular momentum of the pair solution, we obtain
\begin{eqnarray*}
\Phi^s= \frac{1}{\sqrt{6}}
 &&\bigg\{\phi_1(2-3) \; \biggl[ \phi_2(3-1)\; \phi_3(1-2)
            -(-1)^{M_L} \phi_2(1-2)\; \phi_3(3-1) \biggr] \\
 &&+\phi_2(2-3) \; \biggl[ \phi_3(3-1)\; \phi_1(1-2)
            -(-1)^{M_L} \phi_3(1-2)\; \phi_1(3-2) \biggr]\\
 &&+\phi_3(2-3) \; \biggl[ \phi_1(3-1)\; \phi_2(1-2)
            -(-1)^{M_L} \phi_1(1-2)\; \phi_2(3-1) \biggr]
                            \bigg\}
\end{eqnarray*}
where $M_L=\sum_{i=1}^3 m_i$ is the total orbital angular momentum.
If we define a matrix of pair functions
\begin{equation}
{\bf S}=
\left[ \begin{array}{ccc} \phi_1(2-3) & \phi_1(3-1) & \phi_1(1-2)\\
                          \phi_2(2-3) & \phi_2(3-1) & \phi_2(1-2)\\
                          \phi_3(2-3) & \phi_3(3-1) & \phi_3(1-2)
       \end{array} \right]
\end{equation}
then our result can be written as a determinant or 
permanent\footnote{A permanent
 is similar to a determinant without the factors $(-1)^p$}
 of pair functions
\begin{eqnarray}
\Phi^s &=&\frac{1}{\sqrt{6}}\; 
Det({\bf S})~~~~~~~~~~~ {\mbox for} \;\; M_L=even \\
\Phi^s &=&\frac{1}{\sqrt{6}}\; Perm({\bf S})~~~~~~~~ {\mbox for} \;\; M_L=odd
\end{eqnarray}
From this fact we conclude the following important rule
valid for ${\bf R}={\bf 0}$:\\
{\em If at least two pair functions agree, $M_L$ has to be odd.}\\
Otherwise the determinant vanishes. This means in particular, that
the total ground state cannot be built up from identical
ground state pair functions
with $m=0$, which would have the lowest energy without taking
the Pauli principle into account.
Therefore, the ground state
consists of one pair with $m=1$ and two pairs with $m=0$ and
has a total angular momentum $M_L=1$ in the strong correlation limit.\\
Observe the difference of these conclusions to the Pauli principle for one--
electron orbitals. If $\Phi$ were a product of functions
which  depend on one coordinate only (one--particle orbitals),
which is equivalent to a Slater determinant for  $\Phi^s$,
in the quartet state all three spatial functions must be different, and the
ground state would have a total angular momentum
$M_L=(m-1)+m+(m+1)=3 m$ where the integer $m$ depends on the strength
of the magnetic field. This holds if $\tilde\omega$ is so large that
no {\em excited} one--particle orbitals are involved, i.e. in the
 weak correlation regime .
Indeed,  for strong fields the simple rule that $M_L$ is a multiple
of 3 is confirmed also by numerical
calculations \cite{Girvin},\cite{Maksym},\cite{Hawrylak}.\\

{\bf Doublet $S=\frac{1}{2}$}\\
Because of $f=2$ there are two degenerate and orthogonal
functions $\Psi_i$ which span a
subspace. This level does not exist
from symmetry reasons only if both functions vanish.
We assume ${\bf R}={\bf 0}$ so that (\ref{inversion}) holds and
consider the two
vectors $\bf \Phi \mbox{$^s_i$}$ one by one. \\
For $i=1$ and $M_L=even$ we obtain
\begin{eqnarray*}
{\bf \Phi \mbox{$^s_1$}}=\frac{1}{\sqrt{3}}                      \bigg\{ \;\;\;
\left[ \begin{array}{c} 1\\0 \end{array} \right] \;
               \phi_3(1-2) \;\; D_{12}(2-3,3-1) &&\\
+ \frac{1}{2} \left[ \begin{array}{c} -1\\-\sqrt{3} \end{array}  \right] \;
               \phi_3(2-3) \;\; D_{12}(3-1,1-2)  &&\\
+ \frac{1}{2} \left[ \begin{array}{c} -1\\ \sqrt{3} \end{array} \right]  \;
               \phi_3(3-1) \;\; D_{12}(1-2,2-3)  &&        \bigg\}
\label{i1}
\end{eqnarray*}
where
\begin{equation}
D_{12}(2-3,3-1)=
Det \; \left[ \begin{array}{cc} \phi_1(2-3) & \phi_1(3-1) \\
                          \phi_2(2-3) & \phi_2(3-1) \\
       \end{array} \right]
\label{det}
\end{equation}
and the other determinants are defined analogously: the subscripts refer to the
quantum numbers of the pair functions involved and the arguments define
the arguments of the pair functions. \\
For $i=1$ and $M_L=odd$ we obtain a similar formula as before but
with the {\em determinants} in (\ref{det}) replaced by  {\em permanents}
called $P_{12}$.\\
For  $i=2$ and $M_L=odd$ the result is also similar
 but with different column vectors\\
\begin{eqnarray*}
{\bf \Phi \mbox{$^s_2$}}=\frac{1}{\sqrt{3}}                      \bigg\{ \;\;\;
\left[ \begin{array}{c} 0\\1 \end{array} \right] \;
               \phi_3(1-2) \;\; D_{12}(2-3,3-1)&& \\
+ \frac{1}{2} \left[ \begin{array}{c} \sqrt{3}\\-1 \end{array}  \right] \;
               \phi_3(2-3) \;\; D_{12}(3-1,1-2) && \\
+ \frac{1}{2} \left[ \begin{array}{c} -\sqrt{3}\\ -1 \end{array} \right]  \;
               \phi_3(3-1) \;\; D_{12}(1-2,2-3) &&         \bigg\}
\label{i2}
\end{eqnarray*}
where $D_{12}$ is defined as in (\ref{det}).\\
For $i=2$ and $M_L=even$ the {\em determinants} in the last formula
 have to be replaced  by  {\em permanents}.\\

It is obvious that for $\phi_1=\phi_2 (\neq \phi_3)$  the determinants
$D_{12}$ vanish and consequently for
$M_L=even$, $\bf \Phi \mbox{$_1^s$}$ vanishes and for
$M_L=odd$, $\bf \Phi \mbox{$_2^s$}$ vanishes, but both do not
vanish simultaneously.
Therefore this state is allowed.
However, if all three pair functions agree, also those functions vanish in which
permanents occur, because the sum of the prefactors sum up to zero.
Consequently, the Pauli--principle tells us that for ${\bf R}={\bf 0}$
in the doublet state:\\
{\em All three pair functions must not agree.}\\
 This is analogous to the
one--particle model, where two orbitals may agree
 (if they occupy different spin states), but not all three of them.
This means that, as for the quartet,
the ground state consists of one pair with $m=1$ and two pairs with $m=0$ and
it has a total angular momentum $M_L=1$. Thus, in our limit and for vanishing
magnetic field, the quartet and the doublet ground state energy agree.
This degeneracy is lifted by a magnetic field because of the Zemann energy.

\section{Acknoledgement}

I am indebted to  H.Eschrig, W.Weller and their groups as well as
J.K.Jain,  W.Apel, and U.Z\"ulicke  for discussion and the
Deutsche Forschungs-- Gemeinschaft
for financial funding.

\end{document}